\documentclass[journal=jpccck,manuscript=article]{achemso}

\usepackage{amssymb}
\usepackage{indentfirst}
\usepackage{chemformula} 
\usepackage[T1]{fontenc} 
\usepackage{chemformula} 
\usepackage[T1]{fontenc} 
\usepackage{amssymb}
\usepackage{indentfirst}
\usepackage{chemformula} 
\usepackage[T1]{fontenc} 
\usepackage{textgreek}
\usepackage[version=3]{mhchem} 
\usepackage{xr}
\usepackage{booktabs}
\usepackage{xcolor}
\usepackage{epstopdf}
\usepackage{amsmath} 
\newcommand{\angstrom}{\text{\normalfont\AA}}
\AppendGraphicsExtensions{.tga}
\usepackage[labelformat=parens,labelsep=quad,skip=3pt]{caption}
\usepackage{graphicx}
\usepackage{hyperref}
\usepackage{caption}
\usepackage{hyphenat}

\author{Zekun Chen}
\author{Fernanda C. Bononi}
\author{Charles A. Sievers}
\author{Wang-Yeuk Kong}
\author{Davide Donadio}
\email{ddonadio@ucdavis.edu}
\affiliation[UCDChem]
{Department of Chemistry, University of California Davis}

\title{UV-Visible Absorption Spectra of Solvated Molecules by Quantum Chemical Machine Learning}

\begin{document}

\begin{tocentry}
\begin{center}
    \includegraphics[width = 1\linewidth]{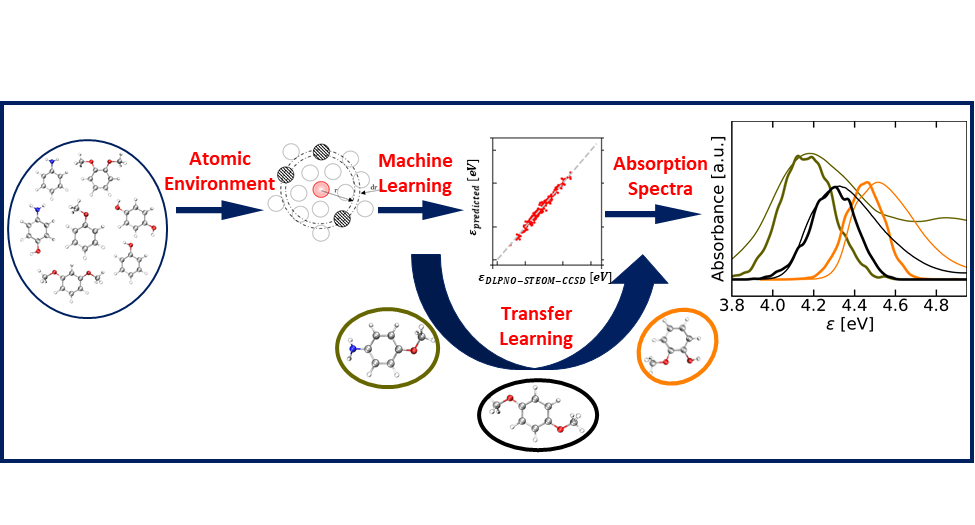}
\end{center}
\end{tocentry}


\begin{abstract}
Predicting UV-visible absorption spectra is essential to understand photochemical processes and design energy materials. Quantum chemical methods can deliver accurate calculations of UV-visible absorption spectra, but they are computationally expensive, especially for large systems or when one computes line shapes from thermal averages. Here, we present an approach to predict UV-visible absorption spectra of solvated aromatic molecules by quantum chemistry (QC) and machine learning (ML). We show that a ML model, trained on the high-level QC calculation of the excitation energy of a set of aromatic molecules, can accurately predict the line shape of the lowest-energy UV-visible absorption band of several related molecules with less than 0.1 eV deviation with respect to reference experimental spectra. Applying linear decomposition analysis on the excitation energies, we unveil that our ML models probe vertical excitations of these aromatic molecules primarily by learning the atomic environment of their phenyl rings, which align with the physical origin of the $\pi\rightarrow\pi^\star$  electronic transition. Our study provides an effective workflow that combine ML with quantum chemical methods to accelerate the calculations of UV-visible absorption spectra for various molecular systems.

\end{abstract}

\section{Introduction}
Improving our understanding and ability to model light-matter interactions is essential to several branches of chemical research, including biochemistry, environmental chemistry,\cite{domine_air-snow_2002} and renewable energy harvesting conversion.
The accurate prediction of UV-visible light absorption spectra is often the first step in the design molecular chromophores, light-harvesting complexes, organic photovoltaics, dyes for photoelectrochemical cells, photoresponsive materials, photocatalysts, and food dyes.\cite{pastore2010computational,yu_recent_2015,romero2017quantum, bertrand2017photo, hullar2020photodecay,  denish_discovery_2021,karuthedath_intrinsic_2021} 

Many-body quantum chemical approaches, such as the Bethe-Salpeter equation\cite{blase2018bethe}, linear response (LR) or equation of motion (EOM) coupled-cluster (CC) theories\cite{dreuw2005single} provide excitation energies with accuracy that approaches the golden standard reference of full configuration interaction calculations.\cite{christiansen_excitation_1996,veril2021questdb}.
However, the steep computational cost scaling of these methods (e.g. $O(N^{7})$ for LR-CC3, and $O(N^{6})$ for EOM-CCSD, i.e. EOM-CC with singles and doubles)\cite{loos2020quest} makes their applications to large molecular systems impractical. 
Moreover, to accurately predict spectral line shapes at finite temperature, it is necessary to perform statistical averages over several hundred or even thousands of configurations, thus further increasing the computational demand.\cite{barone2010theoretical,Malcioglu,de2013absorption,ge_accurate_2015, timrov2016multimodel,zuehlsdorff2018combining, zuehlsdorff2018unraveling, zuehlsdorff2019optical, bononi2020bathochromic,feher2021multiscale} 

Machine Learning (ML) is emerging as an invaluable tool to accelerate quantum chemistry (QC) calculations, providing accurate results at a fractional cost of electronic structure calculations.\cite{butler_machine_2018,westermayr2021perspective} \textcolor{black}{Former studies applied ML to model electronic excitations in molecules inferring structure-property relations from short-range representations of the molecular geometries.\cite{montavon2013machine,ramakrishnan2015electronic,pronobis2018capturing,bartok2013representing,ramprasad2017machine,pozdnyakov2020incompleteness,townsend2020representation, Musil2021}.} However, to the best of our knowledge, very few works have focused on accurately predicting the  line shapes of UV-visible absorption spectra from ML models.
Ye {\it et al.} used an artificial neural network with internal coordinates and Coulomb Matrices as an input layer to fit the electronic absorption spectrum of N-methylacetamide using molecular structures sampled from classical molecular dynamics (MD) trajectories.\cite{rupp2012fast, abiodun2018state, ye2019neural}
Xue {\it et al.} fitted the absorption cross-sections of benzene and 9-Dicyanomethylene using a kernel ridge regression with the displacements from the equilibrium geometry as descriptors.\cite{murphy2012machine,xue2020machine} A similar approach, which employs linear fitting of excitation energies against molecular coordinates, was used to improve the statistical sampling of the UV-visible absorption spectra of phenol and guaiacol at the air-ice interface and in aqueous solutions.\cite{bononi2020bathochromic}
These works rely on time-dependent density functional theory (TDDFT)\cite{runge1984density} calculations of the vertical excitation energy (VEE) for several hundreds of molecular configurations to construct suitable training sets, and fit one ML model for each molecule, in order to enhance the convergence of spectral line shapes obtained via statistical averaging. The use of TDDFT limits the accuracy of the VEE calculations, and fitting an {\it ad hoc} ML model for each molecule limits the \textcolor{black}{generalibility} and the predictivity of these approaches. \textcolor{black}{Westermayr {\sl et al.}\cite{westermayr2020deep} stepped further to model UV-Visible Absorption of $CH_{2}N{H_{2}}^{+}$ and $C_{2}H_{4}$ molecules using deep neural-network (DNN) based on the SchNarc\cite{westermayr2020combining} architecture and was able to generalize the DNN to predict UV-Visible absorption for three other small molecules. However, extending this model may not be straightforward, especially for larger molecules, due to the inherent complexity of DNN.}

Here, we devise a framework that overcomes these limitations and allows one to compute spectral line shapes that are comparable to experimental measurements by coupling high-level QC calculations with a ML model that can be applied to several different molecules.
To this scope, we adopt the bispectrum components (BC), an atomic environment descriptor commonly used to develop ML interatomic potentials,\cite{bartok_gaussian_2010,thompson2015spectral,cusentino2020explicit, zuo2020performance}  as the input to a regularized regression model to predict the UV-visible absorption spectra of a set of ten aromatic molecules in aqueous solution. To pursue chemical accuracy we train the ML model on the EOM-CCSD excitation energies of configurations from first-principles molecular dynamics (FPMD) trajectories.
\textcolor{black}{
We show that a single ML model predicts the shape of the lowest-energy UV-visible absorption bands of a set of similar molecules with an accuracy comparable to experimental measurements. 
Furthermore, by training the ML model on a subset of seven molecules we test its capability to retain its predictivity beyond the training set.}

In the next section, we provide the outline of the multiscale modeling approach used to compute the reference UV-visible absorption spectra, the features of the proposed ML model, and its parameterization. In the following section we present the results obtained by fitting a unified ML model over the full dataset of ten molecules. We then tested how this ML approach would be \textcolor{black}{generalized to predict} the UV-visible spectra of the three molecules left out of the training set. 
Finally, we establish a connection between the ML descriptors and the electronic excitations, so to provide a chemical interpretation of the ML results. A concluding section summarizes the key results and highlights future perspectives. 



\section{Methods and Models}

\begin{figure}[H]
    \centering
    \includegraphics[width = 1\linewidth]{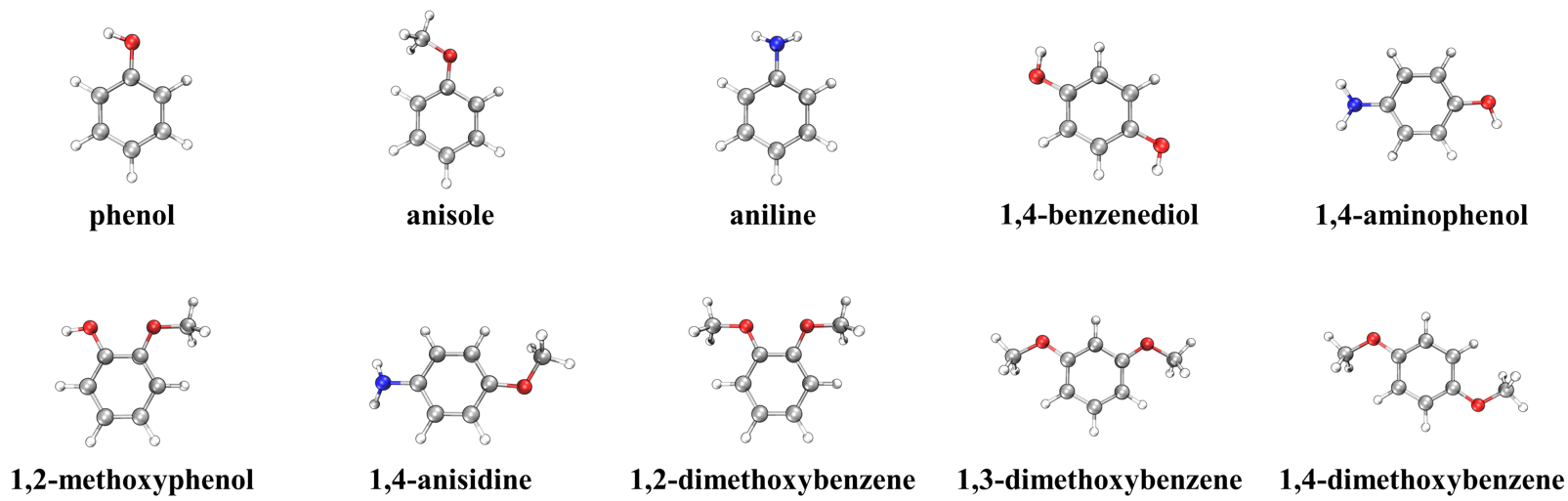}
    \vspace{2 mm}
    \caption{Schematics of the ten aromatic molecules used in this study.}
    \label{fig:molecules}
\end{figure}

Our development of a quantum chemically informed statistical learning method to compute spectra line shapes consists of the following steps: 
\begin{enumerate}
    \item FPMD simulations of molecules in aqueous solutions with explicit water;
    \item Quantum chemical calculation of the (first) excitation energies for a few hundred frames extracted from the FPMD trajectory, in which the explicit solvent is replaced by a polarizable continuum implicit solvation model;
    \item Representation of the molecular geometries in BC;
    \item Fit of a linear ML model of the excitation energies as a function of the molecular configuration by regression with norm-one ($\ell_{1}$) regularization. 
\end{enumerate}
This approach is applied to a set of 10 molecules (Figure~\ref{fig:molecules}) consisting of a benzene ring with different combinations of the following functional groups: --NH$_2$ (amine), --OH (hydroxyl), --OCH$_3$ (methoxy).

\subsection{First-Principles Molecular Dynamics}
The calculation of the UV-visible absorption spectra follows a multiscale approach which combines FPMD and excited state calculation. This approach
accurately predict line shapes as it includes both temperature and solvation effects, within the limits of the method used to compute electronic excitations.\cite{ge_accurate_2015, timrov_self-consistent_2015, timrov2016multimodel, zuehlsdorff_modeling_2019} 
In this work, density functional theory (DFT) based FPMD simulations are performed using the mixed basis-set Quickstep approach, implemented in the CP2K package\cite{vandevondele2005quickstep,hutter2014cp2k}. We use the Perdew-Burke-Ernzerhof (PBE) generalized gradient approximation (GGA) for the exchange and correlation functional\cite{PBE} with D3 van der Waals corrections.\cite{grimme_consistent_2010}
Valence Kohn-Sham orbitals are expanded in real space on a double-$\zeta$ localized valence polarizable (DZVP) basis set\cite{feller1996role}, the electron density in reciprocal space is expanded in plane waves up to a cutoff energy of 300 Ry, and core states are treated implicitly using Geodecker-Teter-Hutter norm-conserving pseudopotentials.\cite{Goedecker} 
We run MD simulations of each molecule in aqueous environment in cubic periodic cells with 128 water molecules. 
FPMD runs were initialized from classical MD simulations in the constant pressure canonical ensemble at ambient conditions, using the TIP4P/Ice model for water and the generalized Amber force field (GAFF) for the organic molecules.\cite{GAFF, abascal2005potential}
These systems were then equilibrated for 10~ps by FPMD at constant volume at 300 K. 
The frames used for absorption spectra calculations were extracted from 100 ps long production runs in the constant-volume canonical ensemble, \textcolor{black}{enforced using the stochastic rescaling algorithm} with a coupling constant of $\tau=10$ ps.\cite{Bussi} The equations of motion were integrated with a timestep of 0.5 fs. 

\subsection{Absorption Spectra Calculation}

For each molecular geometry sampled from the FPMD simulations, VEEs were computed by  “domain-based local pair natural orbital similarity transformed equation of motion coupled cluster singles and doubles” (DLPNO-STEOM-CCSD): an efficient coupled-cluster approach with $O(N{^4})$ scaling.\cite{berraud2019unveiling} 
For the sake of efficiency, the explicit solvent was replaced by a conductor-like polarizable continuum model(CPCM).\cite{pascual1990gepol, silla1991gepol, pascual1994gepol}
All electron DLPNO-STEOM-CCSD calculations were performed with a triple-$\zeta$ \textcolor{black}{valence} polarized Karlsruhe basis set (def2-TZVP)\cite{weigend2005balanced} using ORCA 4.2.1.\cite{neese2020orca}  
For the excited state calculations, the resolution of identity approximation for both Coulomb and exchange integrals was employed to speed up the self-consistent calculation. 
For each molecule, the lowest-energy absorption band of the UV-visible spectrum is simulated by summing Gaussian functions with a width of 0.027 eV centered on the calculated VEE for 200 statistically independent configurations chosen from the FPMD trajectory. \textcolor{black}{This approach corresponds to the ``ensemble method", which provides a reasonable approximation of line widths but does not capture vibronic effects arising from nuclear dynamics.\cite{zuehlsdorff2019optical} Furthermore, we do not re-weight the contributions to the spectrum using the oscillator strength obtained from QC calculations, as we have verified that for each molecule this quantity does not vary significantly over the ensemble of structures that we used to obtain the absorption spectra.}

\subsection{Bispectrum Components}

The goal of our ML model is to predict VEEs as a function of the molecular coordinates, so to spare the computational burden of the QC calculations.  
To this scope, we represent the molecular configuration in BC.\cite{bartok_gaussian_2010} 
While originally BC was proposed as a descriptor to approximate the Born-Oppenheimer potential energy surface of single element systems\cite{bartok_gaussian_2010}, it has been applied to develop ML potential of multi-component systems and predict material properties such as elastic constants, bulk modulus as well as  vibrational free energies and entropies of solids\cite{legrain2017chemical,li2018quantum,cusentino2020explicit}.
Compared to other atomic environment descriptors such as smooth overlap of atomic positions (SOAP)\cite{bartok2013representing, bartok_machine_2017,wilkins_accurate_2019} and atom-centered symmetry functions (ACSF) \cite{behler2011atom,bartok2013representing}, BC is more keen at describing the nuances of atomic environments, as it is projected to a more complete set of basis functions with higher dimensions \cite{pozdnyakov2020incompleteness,zuo2020performance}. 
Additionally, in the development of spectral neighbor analysis potentials, BC was formulated to retain a linear relation with the target property.\cite{thompson2015spectral,cusentino2020explicit}
This development ensures the resulting ML models to achieve robust performance using only a moderate amount of training data.  
Such a trade-off between model complexity and size of the training data is the most critical factor for us to bridge BC with linear regression. 

Hereafter, we summarize the key formulations of BC to supplement the following parameterization of our ML model.
As reported in the original works,\cite{thompson2015spectral} the atomic environment is expressed as the weighted atomic density ($\rho_{i}(r)$):  
\begin{equation}
 \rho_{i}(r) = \delta(r) + \sum_{r_{ii^{'}} < R_{cut,ii^{'}}}f_{cut}(r_{ii^{'}}) \omega_{i^{'}}\delta (r-r_{ii^{'}}),  
\end{equation}
where $r_{ii^{'}}$ is the interatomic distance between the central atom $i$ and the neighboring atom $i^{'}$, $f_{cut}$ is the cutoff function to smoothly decay the neighboring atomic density to 0 at the pair-wised cutoff radius ($R_{cut,ii^{'}}$). $R_{cut,ii^{'}}$ is computed by summing over cutoff radii pairs between central and neighboring atoms.  The dimensionless weighting factor ($\omega_{i^{'}}$) is used to differentiate the neighboring atoms. After expressed as a sum of $\delta$ functions, $\rho_{i}(r)$  is expanded in hyperspherical harmonics ($U_{j} (\theta,\phi,\theta_{0})$):
\begin{equation}
\label{eqn:pho_i}
 \rho_{i}(r) = \sum_{j = 0, \frac{1}{2},1,...}^{\infty} u_{j} \cdot U_{j}(\theta,\phi,\theta_{0}).
\end{equation}
In equation (\ref{eqn:pho_i}), $u_{j}$ is the Fourier expansion coefficient given by the inner product between $\rho_{i} (r)$  and $U_{j} (\theta,\phi,\theta_{0})$. \textcolor{black}{In this implementation}, $j$, $j_{1}$ and $j_{2}$ are truncated so that $j,j_{1},j_{2} \le j_{max}$ to ensure a finite spatial resolution of the weighted atomic density. $2j_{max}$, the even integrable of $j_{max}$,  represents a hyperparameter to dictate the number of BC used to fit the ML model.  With $u_{j}$, the bispectrum components ($B_{j_{1},j_{2},j}$) can be defined as:
\begin{equation}
B_{j_{1},j_{2},j} = \frac{1}{2j + 1} u_{j_{1}}\otimes_{j_{1}j_{2}j} u_{j_{2}}\cdot(u_{j})^{*},  
\end{equation}
where $\otimes_{j_{1}j_{2}j}$ represents a coupling product analogous to angular momentum coupling of spherical harmonics. The $\frac{1}{2j+1}$ prefactor ensures $B_{j,j_{1},j_{2}}$ invariant under permutation of the atom indices. In this work, BC were computed using the FitSNAP package.\cite{github}


\subsection{Machine Learning Model} 

The goal of our ML model is to predict the first VEE from the molecular geometry with quantum chemical accuracy. For this purpose we train a linear model on the DLPNO-STEOM-CCSD/def2-TZVP excited state calculations described above using BC as descriptors of the molecular configurations.
\textcolor{black}{As described in the previous section and in former works,\cite{bartok_gaussian_2010} the BC descriptor consists of projecting the local atomic environment on hyperspherical harmonics for each atomic species.}  
As the BC descriptor is high-dimensional, we apply the least absolute shrinkage and selection operator (LASSO)\cite{Tibshirani:2011ec, scikit-learn}, a linear regression model with norm one ($\ell_{1}$) regularization. Training LASSO consists of minimizing the loss function with respect to the set of coefficients $\beta$: 
\begin{equation}
\Delta_{loss}(\beta)=  \frac{1}{2N}||\varepsilon_{ML} - \varepsilon_{QC}||_{2}^{2} + \alpha||\beta||_{1}, 
\end{equation}
where $N$ is the number of molecular configurations extracted from the FPMD simulations, $\alpha$ is the regularization parameter, and $\varepsilon_{QC}$ are the first excitation energies obtained from the QC calculations.  
The performance of the ML model is assessed through two statistical metrics: mean absolute error (MAE)  and mean signed error (MSE). The latter, defined as:
\begin{equation}
 MSE = \frac{1}{N}\sum_{i = 1}^{N}(\varepsilon_{QC,i} -\varepsilon_{ML, i}),  
\end{equation}
is useful to spot the occurrence of systematic errors in the predicted VEEs.
After training the LASSO model, we compute the UV-visible absorption spectra by estimating $\varepsilon_{ML}$ for 5000 statistically independent configurations from the FPMD trajectories. The final ML spectrum consists of the envelop of Gaussian functions centered on the $\varepsilon_{ML}$ with a width of 0.014 eV. 
\textcolor{black}{
Whereas our ML model is developed in close analogy with the approach used to construct SNAP interatomic potentials,\cite{cusentino2020explicit} the use of LASSO marks the main difference between SNAP and our excited state model. In a forthcoming section, we discuss extensively the importance of using a norm $\ell_1$ selection operator to shrink the space of ML parameters and how it enhances the predictivity of the ML models beyond the set of molecules on which it is trained. 
}

A further advantage of using atomic descriptors and the LASSO model is that we can compute the relative contribution of each atom or group of atoms to $\varepsilon_{ML}$, through a linear decomposition analysis, such that:
\begin{equation}
\varepsilon_{ML} = \varepsilon_{0} + \sum_{i = 1}^{N_{atoms}}\sum_{k = j, j_{1}, j_{2}}^{j_{max}}\beta_{k}^{\gamma_{i}} B_{k}^{\gamma_{i}} = \varepsilon_{0} +  \varepsilon_{ \gamma}, 
\end{equation}
where $\varepsilon_{0}$ is the intercept of the LASSO model and $\varepsilon_{\gamma}$ represents prediction contributions from the atom type $\gamma$ for atom $i$. 
As $\varepsilon_{ML}$ is a scalar quantity and the LASSO model is linear, $\varepsilon_{ML}$ can be partitioned with respect to different functional groups:
\begin{equation}
\varepsilon_{ML} = \varepsilon_{0} + \varepsilon_{NH_{2}} + \varepsilon_{OH} + \varepsilon_{OCH_{3}} + \varepsilon_{C_{6}H_{n}} 
\end{equation}
\begin{equation}
\%_{group} = \frac{(\varepsilon_{NH_{2}} + \varepsilon_{OH} + \varepsilon_{OCH_{3}} + \varepsilon_{C_{6}H_{n}} )}{\varepsilon_{ML} - \varepsilon_{0}}
\end{equation}
where $\varepsilon_{NH_{2}}$, $\varepsilon_{OH}$ and $\varepsilon_{ OCH_{3}}$ are the prediction contributions from the amine, hydroxyl and methoxy groups.  $\varepsilon_{ C_{6}H_{n}}$ is the prediction contributions from carbon and hydrogen atoms within a phenyl ring ($n = 5$ for phenol, anisole, aniline and $n=4$ for 1,2-methoxyphenol, 1,4-aminophenol, 1,4-anisdine, 1,4-benzenediol and dimethoxybenzene isomers). $\%_{group} $ is used to express the prediction contributions by percentage.

\subsection{Parameterization of the Machine Learning Model}

First, to show the baseline performance of our ML approach, we fitted a model using the full data set of ten molecules with 200 VEE per molecule. Then, to test \textcolor{black}{model generalizability}, we developed a 7-molecule model by leaving out of the QC calculations of 1,2-methoxyphenol, 1,4-aminophenol and 1,4-dimethoxybenzene from the training set.
Before optimizing the 7- and 10-molecule models, we set the parameters of the BC descriptor, specifically the weights ($\omega_{atom}$) and the cutoff radius for each atomic species 
($R_{cut, atom}$). 
To optimize $R_{cut, atom}$, we fixed $R_{cut, H}=1.2$~\AA, and performed a grid search on $R_{cut}$ for each heavy atom between 2.6 and 3.2~\AA.
We simplified the choice of $R_{cut,atom}$ parameter by setting $R_{cut, C} = R_{cut, N} = R_{cut}$  and $R_{cut, O} = 1.05R_{cut}$. Since carbon is the major building block of these aromatic molecules, we set the weighting factor of carbon ($\omega_{C}$)  to unity. For the remaining weighting factors ($\omega_{H}, \omega_{O} \ \&\  \omega_{N}$), a constraint of $\omega_{H} < \omega_{N} \le  \omega_{O}$ was imposed. 
After determining $\omega_{atom}$ and $R_{cut,atom}$, we chose $2j_{max}$ based on the number of available training samples. At $2j_{max} = 18$, the total number of BC almost equals the number of training samples ($N_{train}$) used to develop the 7-molecule model. Thus, to retain a similar number of input features and training samples, we chose $2j_{max}=18$. 
\textcolor{black}{The hyperparameters $R_{cut}$ and $\omega_{atom}$ are optimized for the 7-molecule models and their value is used also for the 10-molecule model. In this process no information is used from the three molecules left out of the training set. 
}
Table~\ref{table:BC para} summarizes the hyperparameters used to compute BC. 
\begin{table}[H]
    \fontsize{10}{12}\selectfont
	\centering
	\caption{Optimized hyperparameters used to compute BC}
	\vspace{3 mm}
	\label{table:BC para}
	\begin{tabular}{c c c c c c c c }
		 $\omega_{H}$   &  $\omega_{C}$ & $\omega_{N}$ & $\omega_{O}$ &  $R_{cut, H} [\angstrom]$   &  $R_{cut, C} [\angstrom]$ & $R_{cut, N} [\angstrom]$ & $R_{cut, O} [\angstrom]$\\
		[0.5ex]
		\hline
		\hline
		 0.75 & 1.0 & 0.8 & 0.9 & 1.20  & 2.80 & 2.80 & 2.94  \\
		\hline
	\end{tabular}
\end{table}

The $\ell_{1}$ regularization in LASSO is designed to achieve feature elimination\cite{scikit-learn}, so to prevent overfitting. 
From Table~\ref{table:ML para}, we see that for both ML models, the number of non-zero features is less than 25 $\%$ of the total number of input features ($N_{features}$).  
The 7- and 10-molecule models retain similar fractions of non-zero features with respect to the size of the training sets so that the model performance can be compared. 
A 10-fold cross validation was applied to both models to avoid bias from a specific split of training and testing data sets. 
\begin{table}[H]
    \fontsize{10}{12}\selectfont
	\centering
	\caption{Hyperparameters and statistical metrics for the ML models}
	\vspace{3 mm}
	\label{table:ML para}
	\begin{tabular}{c c c c c}
		 Model   &  $N_{features \ne 0}/N_{features}$  & $2j_{max}$ & $\alpha$ & $MAE_{avg \ \& \ std}$ [meV] \\
		[0.5ex]
		\hline
		\hline
		10-molecule & 318/1544 & 18 & 5.8 $\times 10^{-7}$ & 23.05 $\pm$ 3.92 \\
		\hline
		7-molecule & 225/1544 & 18 & 1.2 $\times 10^{-6}$ & 23.33 $\pm$ 4.13 \\
		\hline
	\end{tabular}
\end{table}

\begin{figure}[H]
    \centering
    \includegraphics[width = 1\linewidth]{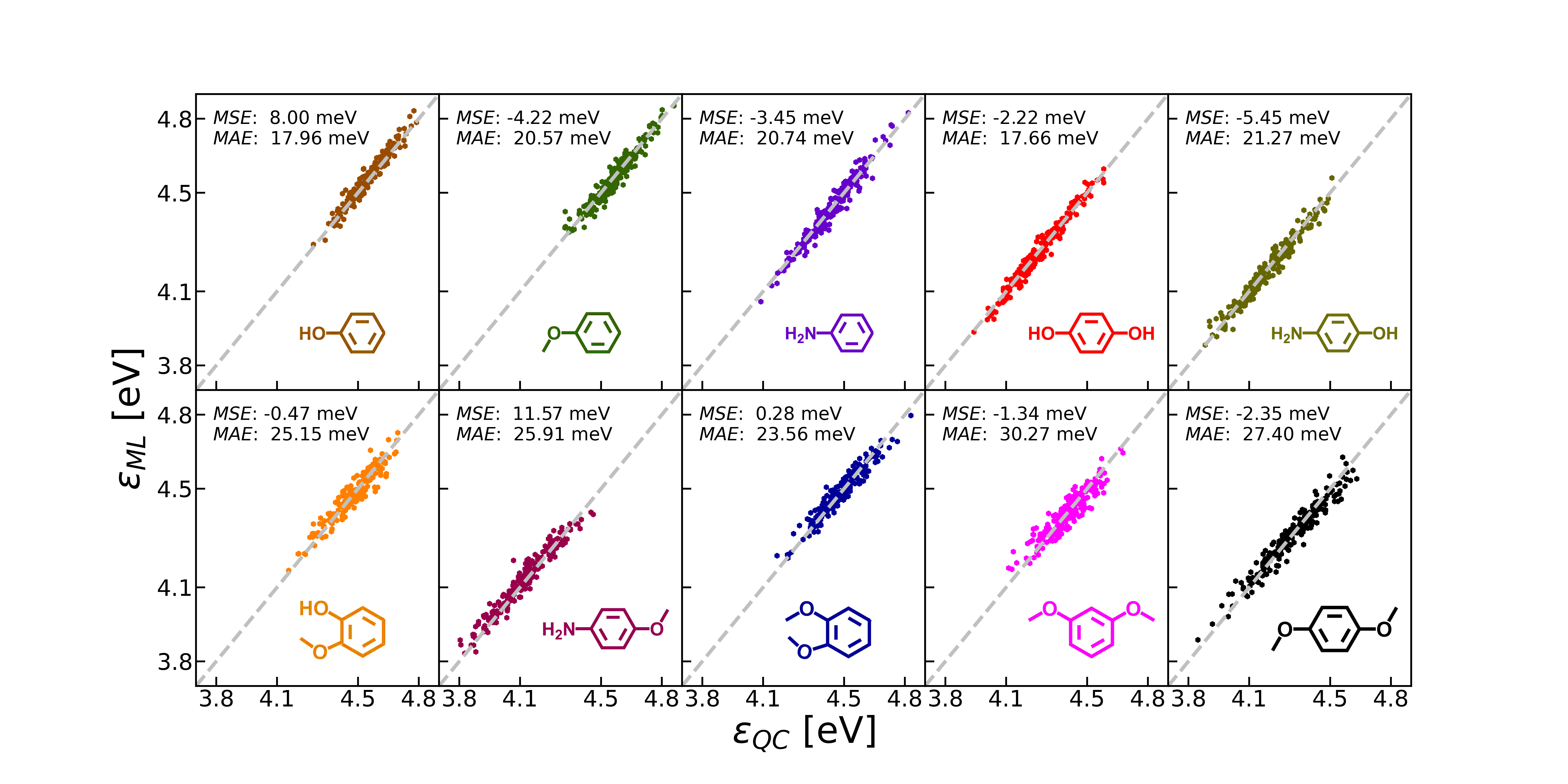}
    \caption{Testing performance for the 10-molecule model. $\varepsilon_{ML}$ is computed by averaging $\varepsilon$ predictions from the 10-fold cross-validation. $\varepsilon_{QC}$ is the quantum mechanically computed excitation energies for the first state.}
    \label{fig:parity_10_mol_model}
\end{figure}

\begin{figure}[H]
    \centering
    \includegraphics[width=1\linewidth]{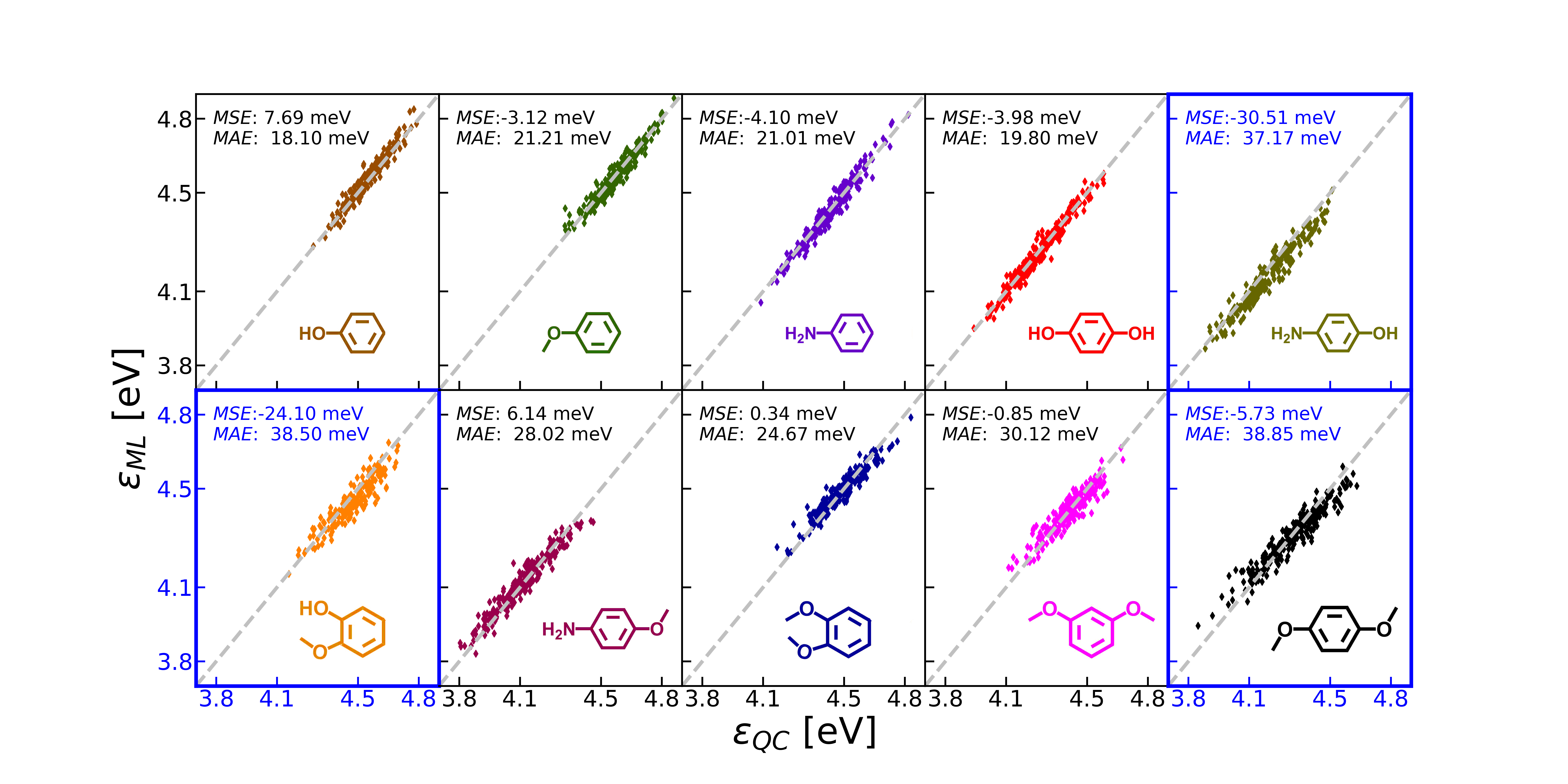}
    \caption{Testing performance for the 7-molecule model. $\varepsilon_{ML}$ is computed by averaging $\varepsilon$ predictions from the 10-fold cross-validation. $\varepsilon_{QC}$ is the quantum mechanically computed excitation energies for the first state. Molecules used in model generalization are indicated with blue panels.}
    \label{fig:parity_7_mol_model}
\end{figure}

As shown in Table \ref{table:ML para},  the testing MAE of our 10-molecule model,  averaged from the 10-fold cross validation, is 23.05 $\pm$ 3.92 meV. A similar MAE, 23.33 $\pm$ 4.13 meV, is also observed from the 7-molecule model. Besides MAE, we also introduced MSE to gauge if our ML models systematically overestimate or underestimate excitation energies. As shown in Figure~\ref{fig:parity_10_mol_model},  most of the molecules have MSE of less than 5 meV, which implies that no noticeable overestimation or underestimation of excitation energies occurs,  \textcolor{black}{except for 1,4-anisidine, for which the predicted excitation energies have MAE and MSE of 25.91 and 11.57 meV.}
Figure~\ref{fig:parity_7_mol_model} illustrates the testing performance of the 7-molecule model. It can be noticed that both 10 and 7-molecule models have similar testing MAE and MSE for molecules in the training set. 
It is worth highlighting, from the 10-molecule model,  that the predicted excitation energies for every single molecule have MAE far below the intrinsic error of the underlying EOM-CCSD method  (70 meV)\cite{dutta2018exploring}. Even for the excitation energies interpolated from the 7-molecule model, the MAE are merely around half of the 70 meV. Hence, no significant error is introduced by both ML models.


\section{Results and Discussion}

\subsection{Machine Learning Absorption Spectra}
\begin{table}[H]
    \fontsize{10}{12}\selectfont
    \centering
	\caption{Summaries of $\varepsilon_{max}$ from experiment, QC and ML spectra. Absolute difference of $\varepsilon_{max}$ between experiment and QC as well as between experiment and ML are indicated in the parenthesis. Molecules used in model-generalization are indicated with blue and bold fonts.}
	\label{table:epsilon_max}
	\begin{tabular}{c c c c c }
		 Molecule &  $\varepsilon_{max,experiment}$ [eV]  & \ $\varepsilon_{max,QC}$ [eV] & \ $\varepsilon_{max,ML_{10-mol}}$ [eV] & \ $\varepsilon_{max,ML_{7-mol}}$ [eV]\\
		[0.5ex]
		\hline
		\hline
		phenol & 4.573  & 4.548 (0.025)  & 4.568 (0.005) &  4.569  (0.004) \\
		anisole & 4.645  & 4.560 (0.085) &  4.553 (0.092) & 4.556 (0.089) \\
		aniline & 4.466  & 4.423 (0.043)  &  4.424 (0.042) &  4.421 (0.045)  \\
		1,4-benzenediol & 4.291  & 4.231 (0.060) & 4.265 (0.026) & 4.260 (0.031)     \\
		\textcolor{black}{\bf{1,4-aminophenol}} & \textcolor{black}{\bf{4.239}}  & \textcolor{black}{\bf{4.163 (0.076)}} & \textcolor{black}{\bf{4.184 (0.059)}} & \textcolor{black}{\bf{4.162 (0.077)}} \\
		\textcolor{black}{\bf{1,2-methoxyphenol}}  & \textcolor{black}{\bf{4.535}} & \textcolor{black}{\bf{4.481 (0.054)}} &  \textcolor{black}{\bf{4.479 (0.056)}} & \textcolor{black}{\bf{4.451 (0.084)}}  \\
		1,4-anisdine  & 4.263 & 4.078 (0.185) & 4.106 (0.157) & 4.106 (0.157)    \\
		1,2-dimethoxybenzene & 4.556  & 4.470 (0.086) & 4.478 (0.078) & 4.479 (0.077) \\
		1,3-dimethoxybenzene & 4.559 & 4.410 (0.149) & 4.408 (0.151)  & 4.409 (0.150) \\
		\textcolor{black}{\bf{1,4-dimethoxybenzene}} &\textcolor{black}{\bf{4.339}} & \textcolor{black}{\bf{4.284 (0.055)}}  & \textcolor{black}{\bf{4.305 (0.034)}}  & \textcolor{black}{\bf{4.301 (0.038)}}\\
		\hline
	\end{tabular}
\end{table}

\begin{figure}[!ht]
    \centering
    \includegraphics[width = 1\linewidth]{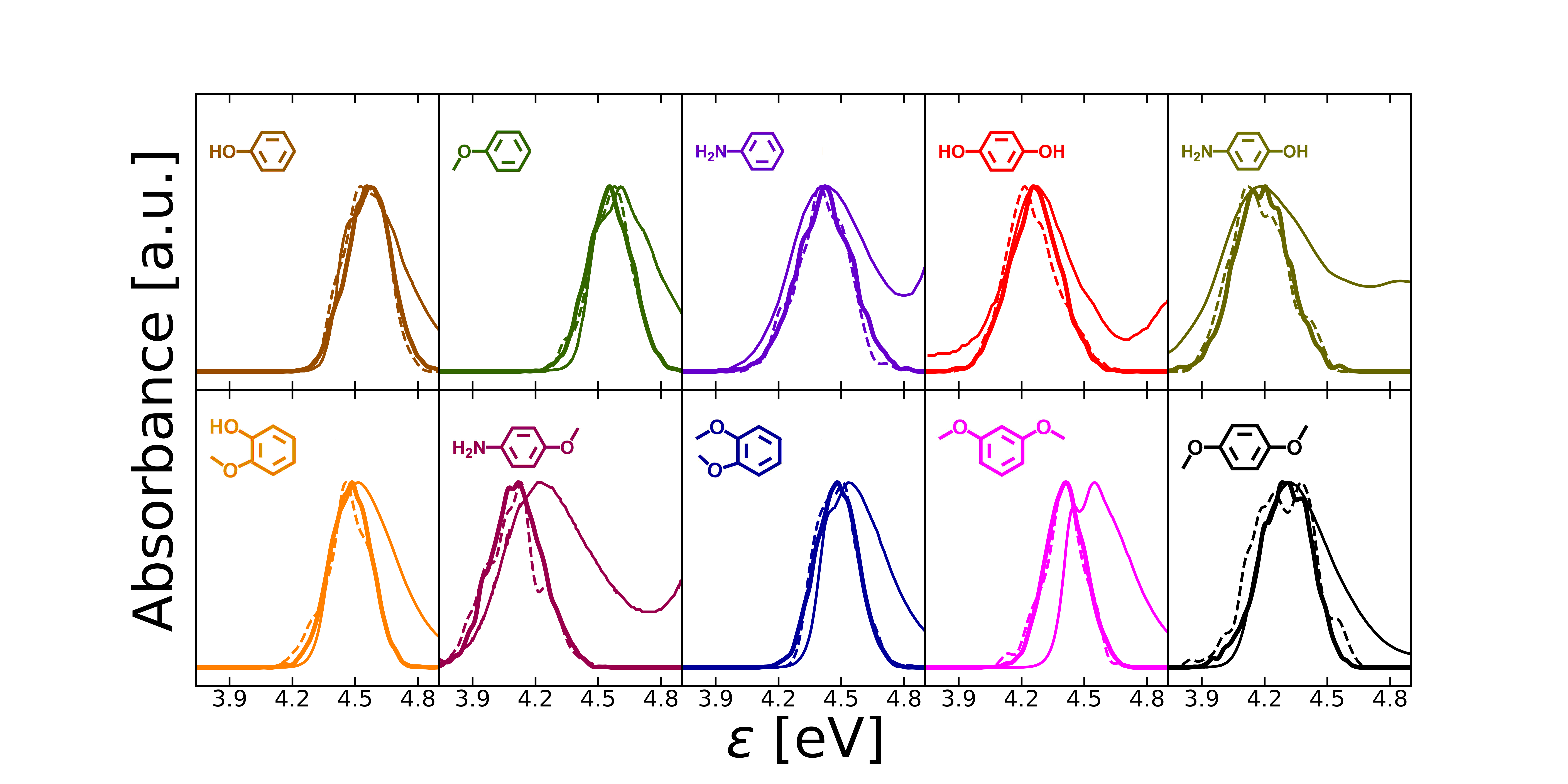}
    \caption{UV-Visible absorption spectra for all 10 aromatic molecules. Thick lines represent the ML spectra predicted using the 10-molecule model and computed with the ensemble method. Thin lines represent the experimental reference\cite{hullar2020photodecay, malongwe2016spectroscopic, chen2020alkaline, corrochano2017photooxidation,stalin2005spectral,shiobara2003substituent}. Dashed lines represent the calculated spectra using the multiscale quantum chemical method.}
    \label{fig:10_mol_model}
\end{figure}

Figure~\ref{fig:10_mol_model} compares the lowest-energy UV-visible absorption band computed by the multiscale quantum-chemical approach described in the methods section to experimental measurements for each molecule in aqueous solution.\cite{shiobara2003substituent,stalin2005spectral,malongwe2016spectroscopic,corrochano2017photooxidation, hullar2020photodecay, chen2020alkaline,hullar2021enhanced}
The QC model is in excellent agreement with experiments for 8 out of 10 molecules, with differences in the center of the peak within less than 0.1 eV and nearly overlapping low-energy tails. Experimental bands tend to be broader on the high energy side, possibly because they encompass the tails of higher excited states.
1,4-anisidine and 1,3-dimethoxybenzene make an exception, with differences in peak positions of 0.18 and 0.15 eV. A possible source of discrepancy is that the GGA functional used in FPMD simulations leads to systematic errors in the geometry of the molecules.
The differences between the gas-phase excitation energies of 1,3-dimethoxybenzene computed for geometries optimized with the PBE and PBE0\cite{adamo1999toward} functionals support this hypothesis (Table S1). \textcolor{black}{However, the cost of running FPMD with PBE0 would be excessive for the purpose of this study. Choosing a semilocal GGA
functional strikes the desired balance between the accuracy of computing spectra and a
reasonable computational cost.} 
Additionally, the experimental conditions may be slightly different from those in the models. The absorption spectrum of 1,4-anisidine was measured at pH$=6.0$, while in our FPMD simulations molecules solvated in aqueous solution at neutral pH, and this difference may cause a shift in the UV-visible absorption.\cite{shiobara2003substituent,zheng2018unexpected} 

The QC spectra were obtained by averaging only a few hundred frames for each molecule. Hence, the calculated spectra for some of these molecules, especially 1,4-dimethoxybenzene, exhibit jagged line shapes (Figure S1). As shown in previous works, ML may be employed to obtain smoother and more refined line shapes.\cite{bononi2020bathochromic,xue2020machine} 
The same effect is achieved here, as illustrated in Figure \ref{fig:10_mol_model}. The ML spectra are obtained by averaging over 5,000 frames for each molecule, which guarantees smooth line shapes and converged statistical sampling. At this point, convergence is limited only by the capability of FPMD simulations to sample the configurational space of the molecules in aqueous solution.
The main features of the ML spectra obtained with this model are almost indistinguishable from the QC references as ML predictions have the same level of  accuracy as the training data.
From Figures \ref{fig:parity_10_mol_model} and \ref{fig:10_mol_model}, we can conclude that a unified ML model is able to accurately predict the excitation energies and thus allowing us to compute the spectra for thousands of molecular configurations at nearly no additional computational cost.

\subsection{Transferability of the Machine Learning Model}

\begin{figure}[!ht]
    \centering
    \includegraphics[width = 1\linewidth]{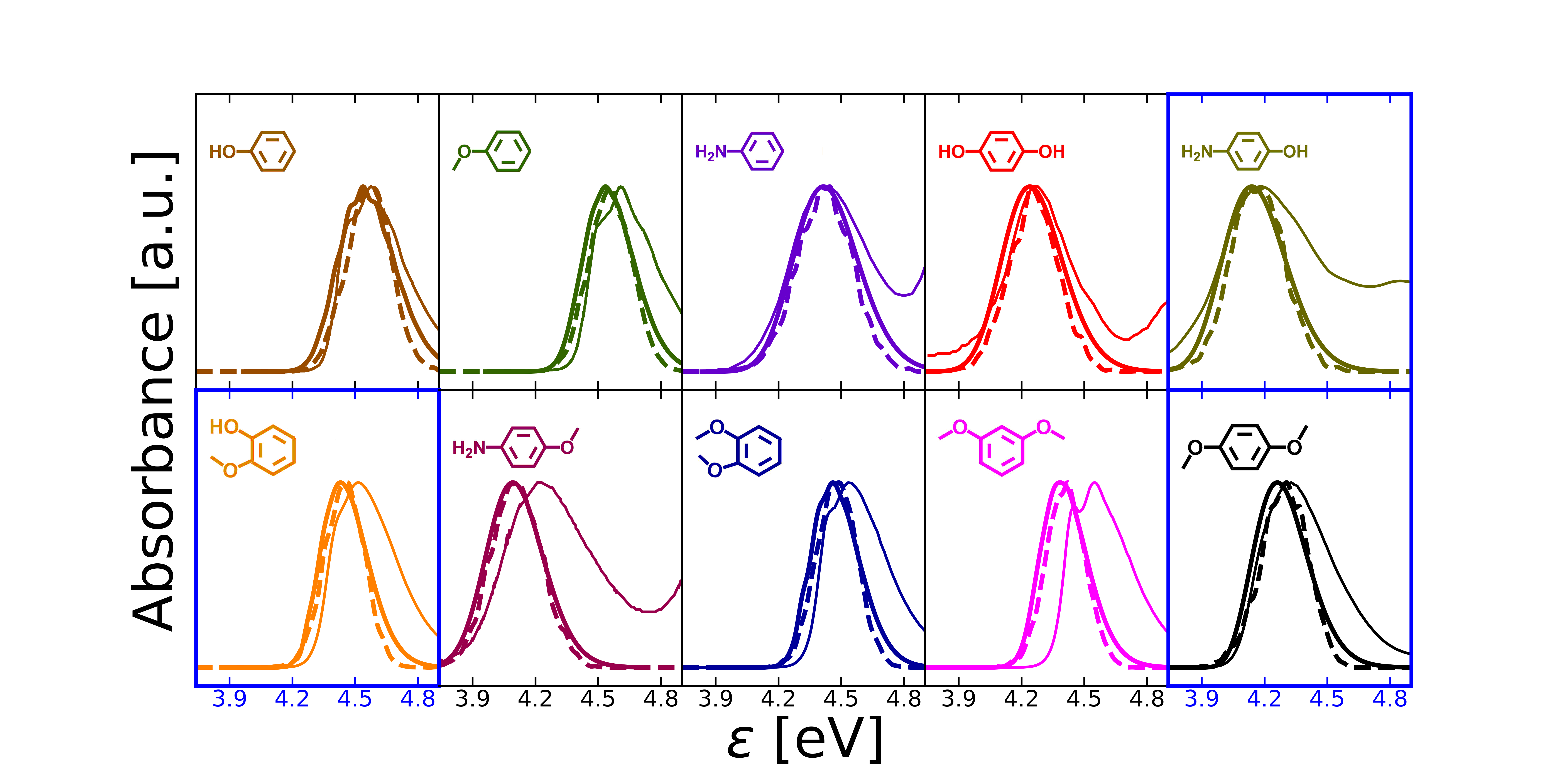}
    \caption{\textcolor{black}{UV-Visible absorption spectra for all 10 aromatic molecules. Thick lines represent the ML spectra computed  with the ensemble method (dashed) and with the third order cumulant scheme (solid).\cite{zuehlsdorff2019optical} Thin lines represent the experimental references\cite{hullar2020photodecay, malongwe2016spectroscopic, chen2020alkaline, corrochano2017photooxidation,stalin2005spectral,shiobara2003substituent}.  
The spectra of the molecules not included in training set are highlighted with blue graph frames and labels.}}
    \label{fig:7_mol_model}
\end{figure}

Although useful to save computational time \textcolor{black}{to compute the lowest energy band of the absorption spectra, the application of the ML model described so far is limited to the molecules comprised in the training set.} 
Hereafter, we explore whether our approach can be used to predict the absorption spectra of molecules that were not included in the training set, \textcolor{black}{provided that these molecules are similar to those used to construct the ML model.}
\textcolor{black}{To this scope, we employ the LASSO model fitted with the excited state calculations of 7 molecules to interpolate the excitation energies and the corresponding UV-Visible absorption spectra for 1,2-methoxyphenol, 1,4-aminophenol and 1,4-dimethoxybenzene. 
These three molecules consist of a phenyl ring with side groups of the same kind (methoxy, hydroxyl and amine) as the molecules included in the training set, but combined to form different isomers or different molecules altogether.}
Table \ref{table:epsilon_max} and Figure~\ref{fig:7_mol_model} show that the peak positions and the absorption line shapes for the seven molecules in the training set remain the same as those obtained with the 10-molecule ML model. 
Figure~\ref{fig:parity_7_mol_model} indicates an increase of both MAE and MSE for the three molecules \textcolor{black}{excluded from the training set}. MAE, in particular, raises beyond 20 meV for 1,4-aminophenol and 1,2-methoxyphenol. 
A larger error for these molecules is expected, 
as the side-group arrangements (e.g., ortho positioning of hydroxyl and methoxyl group, para positioning of amine and hydroxyl group and of two methoxyl groups) are not explicitly known by the ML model and their effects on the excitation energies are inferred from the other seven molecules.
Nevertheless, a close comparison between Figures~\ref{fig:10_mol_model} and \ref{fig:7_mol_model} suggests that the effect of these errors on the predicted UV-visible absorption spectra is small. 
In fact, the model-generalized spectra remain very similar to those obtained with the 10-molecules ML model, are in excellent agreement with the reference QC spectra. 
The difference between $\varepsilon_{max,experiment}$ and $\varepsilon_{max,ML}$ is 0.084 eV,  0.077 eV and 0.038 eV for 1,2-methoxyphenol, 1,4-aminophenol and 1,4-dimethoxybenzene. 
Considering that the 7-molecule model is developed without using any QC calculations of the three excluded molecules, less than 0.1 eV differences are remarkable. 

\textcolor{black}{
These results suggest that the 7-molecule model can properly interpolate the electronic excitations of the molecules not included in the training set without further tuning the hyperparameters of the BC descriptors and provide UV-visible absorption spectra in good agreement with experiments. However, while the low-energy tail of the absorption band is reproduced very well, our approach fails to reproduce the skewed experimental line shapes. To improve the line shape predictions we have applied the dynamics-based third-order cumulant scheme, using the fluctuations of the VEE estimated by the ML model along the FPMD trajectory (Fig.~\ref{fig:7_mol_model} solid line).\cite{mukamel_fluorescence_1985,zuehlsdorff2019optical,chen2020exploiting} 
Whereas we get a minor systematic improvement in the prediction of the line shapes with this approach, the theory still underestimates the intensity of the high-energy part of the absorption band. 
This may be due to the fact that nuclei are treated as classical particles in FPMD, thus neglecting nuclear quantum effects.}

This result shows that local geometric descriptors are sufficient to predict very accurately excitation energies of a group of molecules with similar features. This is a promising starting point for future work on more complex molecules. 
However, to extend the transferability of this approach to other classes of molecules, given the non-local nature of electronic excitations in large molecules, it may turn out necessary to supplement our approach by including richer physically-based descriptors, e.g. electronic orbitals, and to compute excitation energies as the eigenvalues of ML effective Hamiltonians.\cite{zhang2020deep,westermayr_physically_2021,nigam2022equivariant}

\subsection{Effect of the $\ell_{1}$ Regularization}

Hereafter, we analyze the features that make the ML model developed in this work accurate and predictive. 
We first examine the importance of $\ell_{1}$ regularization. To this aim, we built another 7-molecule model using ordinary least square (OLS)\cite{puntanen2013methods} as the underlying ML model. BC for this model were computed using the hyperparameters  as summarized in Table S2. The $2j_{max}$ was chosen to be 8 so that both 7-molecule models have similar $N_{features \ne 0}$. Table S2 shows that both 7-molecule models have very similar overall MAE. The 7-molecule + OLS model even achieves lower standard deviation (std) in overall MAE than the 7-molecule model. Surprisingly, two 7-molecule models show striking differences in MAE with respect to each molecule. From Figure \ref{fig:parity_7_mol_model} and S2, one can see that the 7-molecule model has lower MAE for 5 molecules in the training set. The 7-molecule model also records 8.93, 1.37 and 13.63 meV lower in MAE than the 7-molecule + OLS model when interpolating the excitation energies for 1,2-methoxyphenol, 1,4-aminophenol and 1,4-dimethoxybenzene. The std of MAE from the 7-molecule model (0.724 meV) is only about $12.5 \%$ of the std from the 7-molecule + OLS model (5.76 meV) for the three model-generalized molecules. 
The 7-molecule model is fitted against 225 BC selected from 1544 BC generated with $2j_{max}$ up to 18. \textcolor{black}{As shown in Figure S3, $2j_{max} = 18$ is a sufficiently high order to generate BC with optimal MAE.} Meanwhile, feature elimination schemes, such as $\ell_{1}$ regularization, help capture the essential features of the atomic environments and ensure little to no performance loss as the final model is fitted on a carefully-chosen subset of the initial feature space.\cite{imbalzano2018automatic}. Therefore, these 225 BC are composed of comprehensive descriptions of the atomic environments at resolution as high as $2j_{max} = 18$. However, as no feature elimination is imposed in the 7-molecule + OLS model, its feature space can only be confined at lower order of $2j_{max}$ to prevent overfitting. 
In particular, each of 224 BC generated with $2j_{max} = 8$ is used to fit the 7-molecule + OLS model. Thus, despite almost identical $N_{features \ne 0}$, the difference in the detailed description of the atomic environment eventually leads to a drastic difference in the \textcolor{black}{capability} for the two 7-molecule models. 
\textcolor{black}{Whereas a more standard $\ell_{2}$ regularization may be used, the latter is prone to overfitting when the ratio between the number of samples and features approaches unity. It is therefore very advantageous} to impose the $\ell_{1}$ regularization as it not only prevents overfitting, but it also enables the ML model to have a sufficiently large initial feature space, which is critical for the model to be generalized to systems not included in the training set. 

\subsection{Chemical Interpretation of Machine Learning Results}
\begin{figure}[H]
      \centering
      \includegraphics[width=1\linewidth]{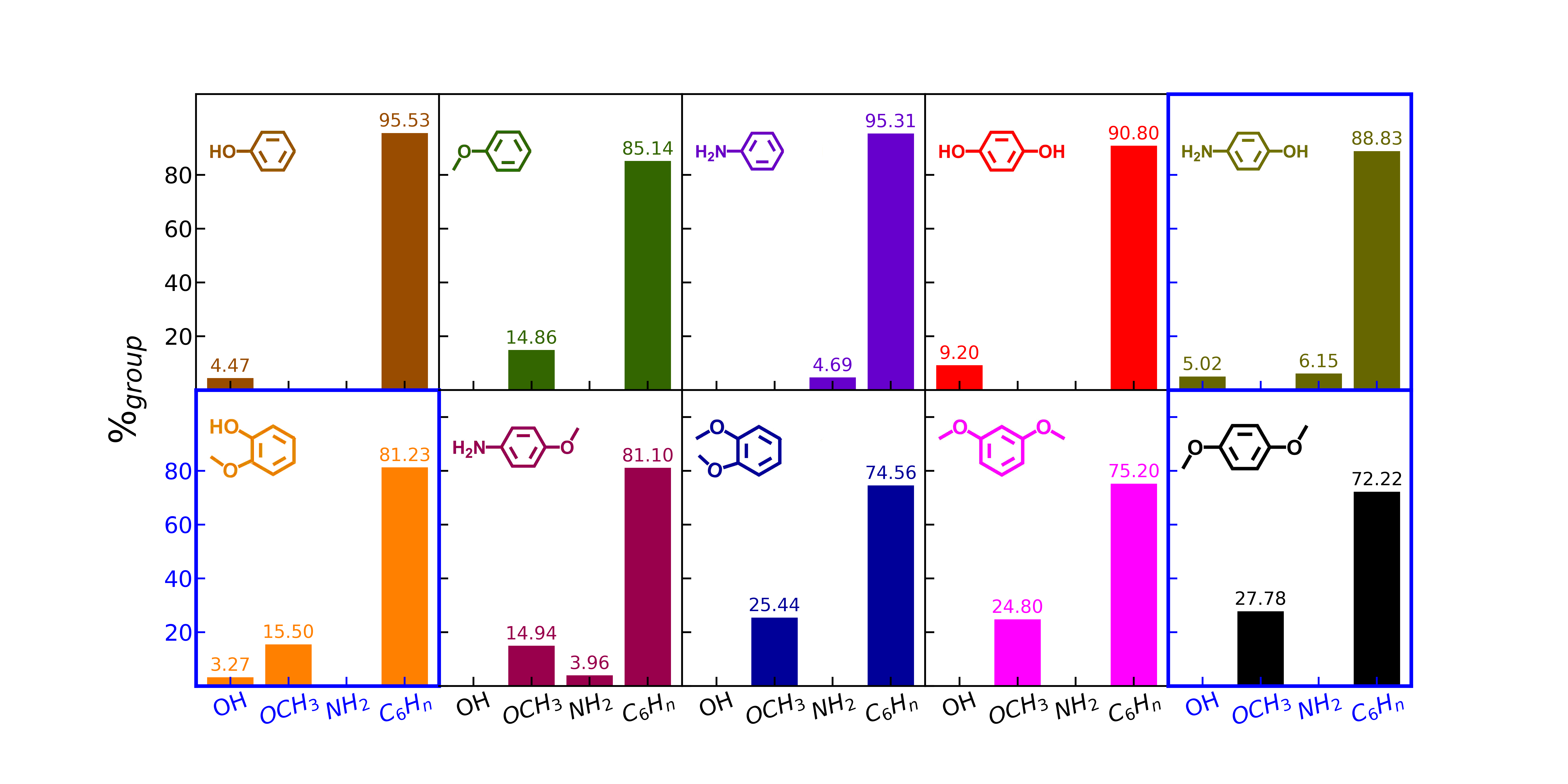}
      \caption{Linear decomposition analysis from the 7-molecule model. $\%_{group}$ are computed by averaging the excitation energy predictions of 5000 frames.}
      \label{fig:7_mol_epsilon}
\end{figure}

To trace how the ML models explore the structural similarity among these aromatic molecules, we performed the linear decomposition analysis as formulated above in the methods section. Figures \ref{fig:7_mol_epsilon} and S4 show that, for $\varepsilon_{ML}$ from both ML models,  contributions from the aromatic ring are \textcolor{black}{predominant}. This trend is common to all molecules. This result suggests that the \textcolor{black}{observed generalizability} across the whole family of aromatic molecules lays primarily on geometric variations of the aromatic ring during the FPMD simulations, and that the primary effects of solvation are local and can be tracked down to configurational changes in the solvated molecule.\cite{p_cooperation_2017}
To interpret this result from a quantum mechanical standpoint, we computed the Natural Transition Orbitals (NTOs)\cite{martin2003natural,badaeva2008two,lu2012multiwfn} of all the aromatic molecules using their equilibrium structures in aqueous solution. \textcolor{black}{As shown in Figures S5, } the dominating NTOs for all these aromatic molecules exhibit the characteristics of a $\pi \rightarrow \pi^{*}$
transition. Such excited state characters are mostly contributed by the aromatic ring, along with a secondary participation of lone pairs on functional groups. 
Whereas relative contributions of the hydroxyl and amine groups from the NTOs are noticeably larger than those illustrated in the linear decomposition analysis, NTOs confirm the predominant relevance of the carbon atoms in the aromatic ring. 
Quantitative differences in the relative contributions of the functional groups indicate that the geometrical interpretation of the excitations from ML is qualitative.

\begin{figure}[H]
    \centering
    \includegraphics[width = 1\linewidth]{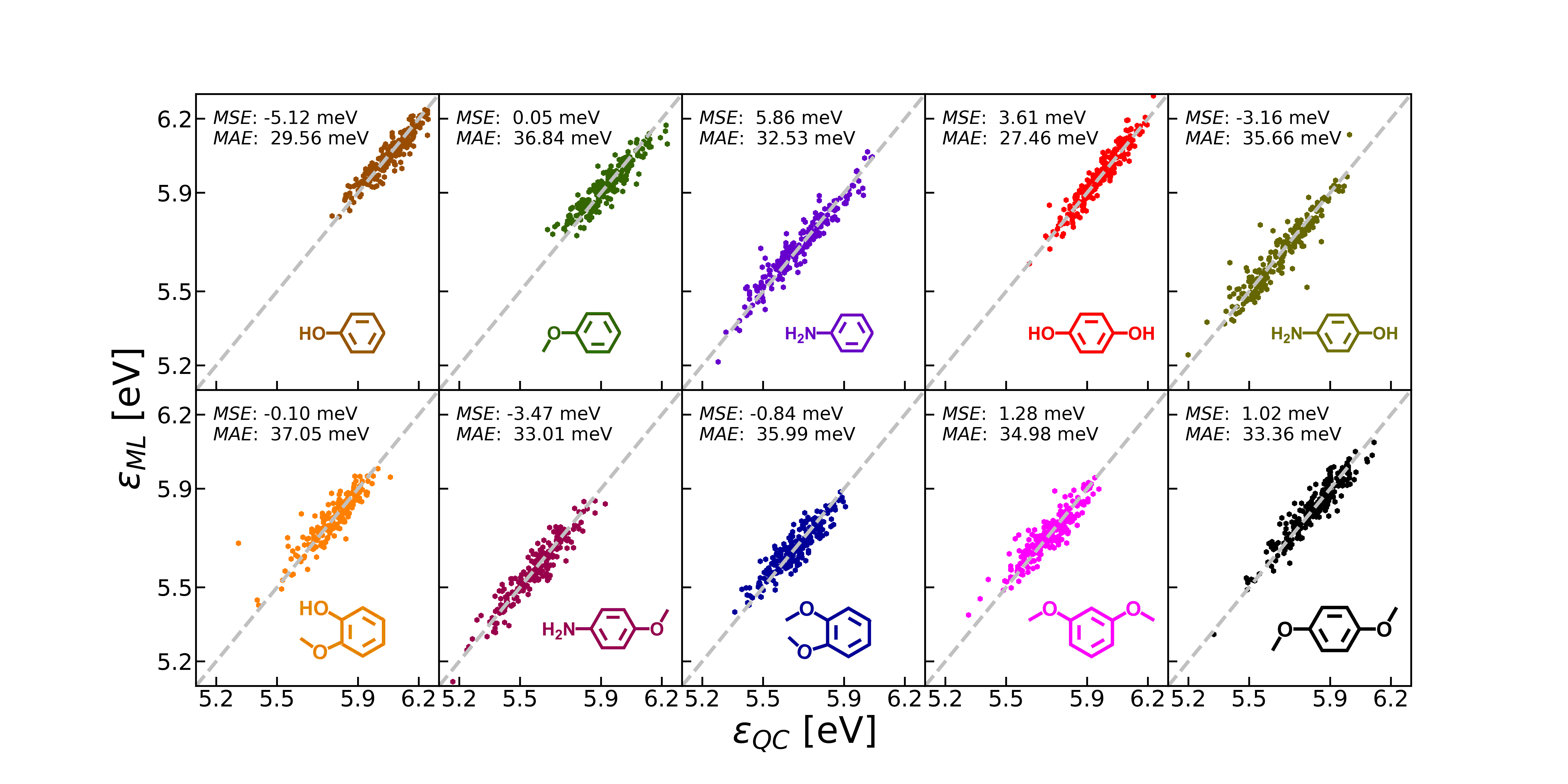}
    \caption{\textcolor{black}{Testing performance for the 10-molecule model. $\varepsilon_{ML}$ is computed by averaging $\varepsilon$ predictions from the 10-fold cross-validation. $\varepsilon_{QC}$ is the quantum mechanically computed excitation energies for the second state.}}
    \label{fig:parity_10_mol_model_2nd_state}
\end{figure}

\begin{figure}[H]
    \centering
    \includegraphics[width = 1\linewidth]{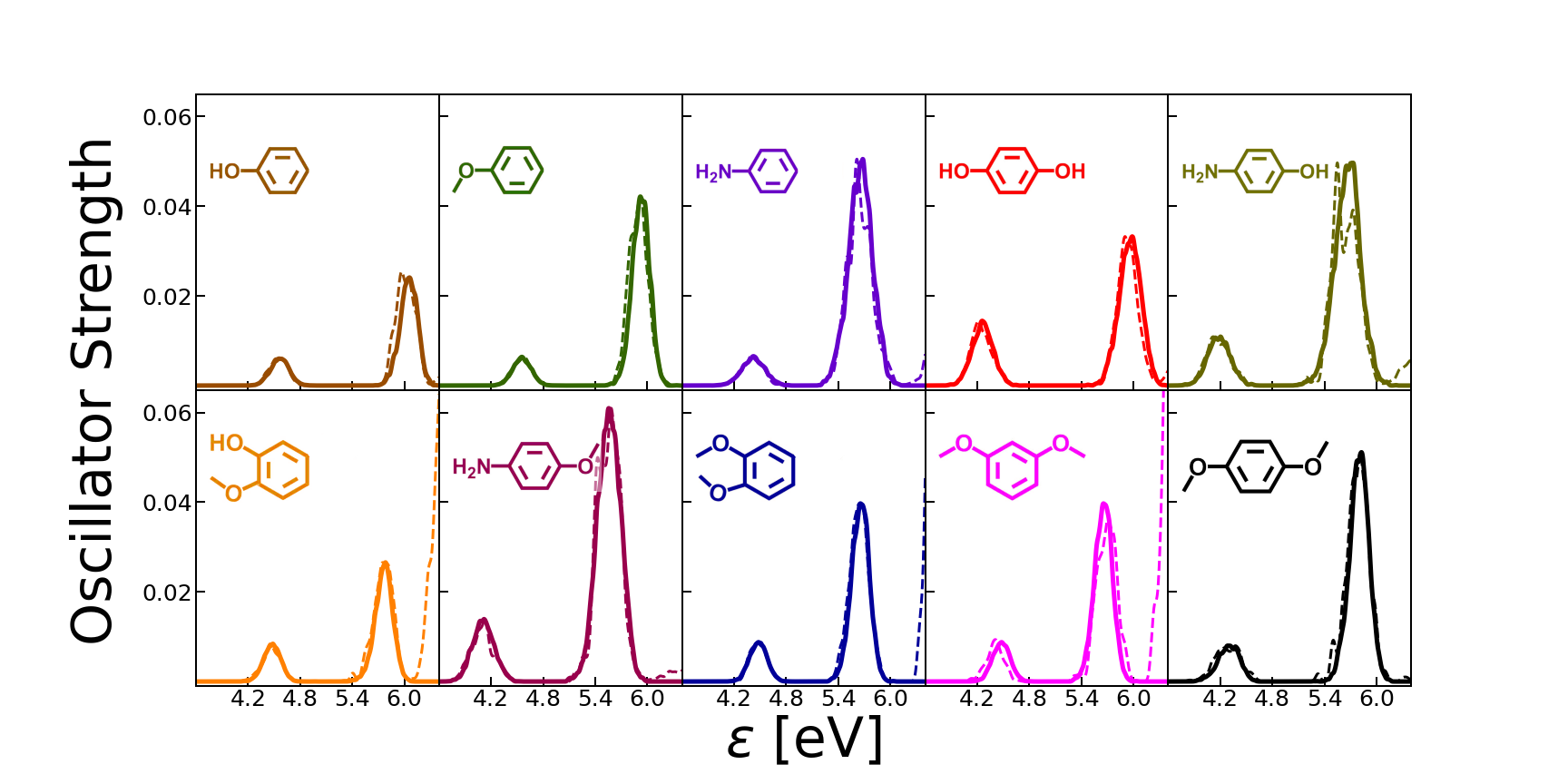}
    \caption{\textcolor{black}{UV-Visible absorption spectra for all 10 aromatic molecules, including the first two excited states. Thick lines correspond to the ML spectra predicted using the 10-molecule model. Dotted lines represent the calculated spectra using the multiscale quantum chemical method.}}
    \label{fig:10_mol_model_2nd_state}
\end{figure}

\textcolor{black}{\subsection{Machine Learning Higher Excited States}}
\textcolor{black}{To explore the possibility of extending our machine learning method, we constructed another model, using the same parameters as the 10-molecule model, to predict the second excitation energies. From Figure~\ref{fig:parity_10_mol_model_2nd_state}, one can see that the MAE and MSE  of the second excited state are noticeably higher than the MAE and MSE  of the first excited state but still far below the 70 meV\cite{dutta2018exploring} limit of the deployed DLPNO-STEOM-CCSD method. With both the first and second excited-state energies predicted concurrently, the resultant ML absorption enrapture long-wavelength tails convoluted from both excited states. Since not all the experimental spectra are available up to second excited states, we compared QC and ML UV-visible absorption bands for the first two excited states in the space of oscillator strength (Figure \ref{fig:10_mol_model_2nd_state}). Both Figures \ref{fig:parity_10_mol_model_2nd_state} and \ref{fig:10_mol_model_2nd_state} prove that our ML methods can be promising in predicting higher excited states. }

\section{Conclusions}
From the ML model performance and the corresponding UV-visible absorption spectra, we have shown that our ML approach can be applied to predict the electronic transitions for a class of solvated aromatic molecules. As a baseline, our 10-molecule model predicts the excitation energies of solvated aromatic molecules with MAE well below the intrinsic error of the underlying QC method. Our 7-molecule model proves that the atomic environment represented by BC can be \textcolor{black}{generalized} to interpolate excitation energies for molecules that are structurally similar to molecules in the training set. By applying our ML models over an ensemble of configurations sampled from FPMD simulations, converged line shapes of the lowest-energy absorption band can be readily attained. The linear decomposition analysis on the predicted excitation energies suggests the aromatic ring to be the key motif to modulate the electronic excitation, which can be explained by the $\pi \rightarrow \pi^{*}$ excited state character of these aromatic molecules. 

This work outlines an efficient strategy to model light absorption spectra for solvated aromatic molecules by combining QC calculations and ML. 
\textcolor{black}{We have shown that, thanks to its modular nature, our workflow can be extended to predict excitation energies of higher-energy states. Since QC-ML allows us to compute efficiently VEE for thousands of frames along a trajectory, we obtained more accurate spectral line shapes combining it with the cumulant expansion method.\cite{chen2020exploiting} Further improved results may be achieved by taking into account nuclear quantum effects in the MD trajectories, e.g., by using path-integral MD and/or a quantum thermostat.\cite{ceriotti_efficient_2010}
}
It would also be possible to attain higher computational efficiency by using accurate $\it ab initio$ potentials, e.g. neural network potentials or extensions of the MBpol model,\cite{schran2021machine, galib_reactive_2021,cruzeiro_highly_2021} instead of FPMD simulations to sample the configurational space of solvated molecules. Besides the computational advantage, some of these models are more accurate than plain DFT with semilocal GGA functionals, as used in this work for FPMD. 
%
In the realm of perspective applications, as the ML model identifies a direct dependence of the excitation energies on the molecular configurations, it would be straightforward to apply this approach to different solvation environments, as shown in the calculation of bathochromic shifts of molecules adsorbed in snow-packs.\cite{hullar2020photodecay,hullar2021enhanced} 
Finally, given the accuracy and the \textcolor{black}{generalizability} of our ML approach, we envisage its extension to broader classes of organic molecules, with potential applications in energy materials, such as organic photovoltaic and dyes for photoelectrochemical systems.\cite{kim2018deep, gupta2021data}.

\section{Acknowledgments}
We are grateful to Ted Hullar, Cort Anastasio and Michele Ceriotti for fruitful discussions on UV-Visible adsorption spectra of aromatic molecules in solution and ML models in computational chemistry. We thank Tim Zuehlsdorff for assisting us with the use of the cumulant scheme code.\footnote{https://github.com/tjz21/Spectroscopy\underline \ python\underline \  code}
We acknowledge support by the National Science Foundation under Grant No. 1806210. Computational resources were provided by the Extreme Science and Engineering Discovery Environment (XSEDE)\cite{XSEDE} (project CHE190009), which is supported by the National Science Foundation, grant number ACI-1548562. 


\begin{suppinfo}
Supporting Information contains excitation energy calculations for the optimized frames of 10 aromatic molecules in gas phase, UV-Vis absorption spectra from  multiscale quantum chemical calculations, comparison summaries of two 7-molecule models, parity plot of 7-molecule + OLS model, linear decomposition analysis for the 10-molecule model as well as Natural Transition Orbitals for all 10 molecules. The data, code and input files that supplement our study are publicly
available at \href{https://github.com/ZKC19940412/mluvspec}{\textcolor{black}{https://github.com/ZKC19940412/mluvspec}}.
\end{suppinfo}

\bibliography{reference}

\end{document}